\documentclass[aps,prd,twocolumn,preprintnumbers,floatfix,reprint,nofootinbib]{revtex4-1}
\usepackage{graphicx}
\usepackage{amsmath}
\usepackage{bm}
\usepackage{yhmath}
\usepackage{subfigure}
\usepackage{color}
\usepackage{cases}
\usepackage{slashed}
\usepackage{times} 
\usepackage{dcolumn,booktabs,bm}
\usepackage{braket}
\usepackage{amsfonts,amssymb,stmaryrd,latexsym}
\usepackage{textcomp}
\usepackage{multirow}
\usepackage{epstopdf}
\usepackage{psfrag}
\usepackage{float}
\usepackage{array} 
\usepackage[colorlinks, citecolor=blue,anchorcolor=red,menucolor=red, linkcolor=red,filecolor=red,runcolor=red,urlcolor=blue,frenchlinks=red]{hyperref}
\graphicspath{{/pic/}}
\newcounter{RomanNumber}

\newcommand{\lyxmathsym}[1]{\ifmmode\begingroup\def\b@ld{bold}
\text{\ifx\math@version\b@ld\bfseries\fi#1}\endgroup\else#1\fi}

\allowdisplaybreaks[4]

\begin{document}

\title{Isospin violating decay $D_s^*\rightarrow D_s\pi^0$ in chiral perturbation theory}

\author{Bin Yang$^{1}$}\email{bin\_yang@pku.edu.cn}
\author{Bo Wang$^{2,1}$}\email{bo-wang@pku.edu.cn}
\author{Lu Meng$^{1}$}\email{lmeng@pku.edu.cn}
\author{Shi-Lin Zhu$^{1,2}$}\email{zhusl@pku.edu.cn}
\affiliation{
$^1$School of Physics and State Key Laboratory of Nuclear Physics and Technology, Peking University, Beijing 100871, China\\
$^2$Center of High Energy Physics, Peking University, Beijing
100871, China}

\begin{abstract}
We systematically calculate the isospin violating decay, $D_s^*\to
D_s\pi^0$, with the heavy meson chiral perturbation theory up to
$\mathcal{O}(p^3)$ including the loop diagrams. The
$\mathcal{O}(p^3)$ tree level amplitudes contain four undetermined
LECs. We use two strategies to estimate them. With the nonanalytic
dominance approximation, we get $\Gamma[D_s^\ast\to
D_s\pi^0]=(3.38\pm0.12)$~eV. With the naturalness assumption, we
give a possible range of the isospin violating decay width,
$[1.11-6.88]$~eV. We find that the contribution of the
$\mathcal{O}(p^3)$ corrections might be significant.
\end{abstract}


\maketitle

\thispagestyle{empty}


\section{INTRODUCTION}\label{sec_intro}

The $D_{(s)}$-mesons are composed of one charm quark and one light
antiquark. The dynamics of $D_{(s)}$-mesons is constrained by both
the chiral symmetry in the light quark sector and the heavy quark
symmetry in the heavy sector. The subtle interplay of the light and
heavy degrees of freedom within the $D_{(s)}$-mesons renders them a
crucial platform to explore and understand QCD. $D_{s0}^\ast(2317)$
and $D_{s1}(2460)$ are two superstars in the $D_s$ family due to
their unexpected low mass. The couple-channel effect between the
$DK^{(*)}$ scattering states and $c\bar{s}$ components leads to the
mass deviation from the quark model
prediction~\cite{Dai:2003yg,Lang:2014yfa,Alexandrou:2019tmk}. See
Ref.~\cite{Chen:2016spr} for a recent review. In addition, the charm
quark mass is not very large. Thus decay behaviors of
$D_{(s)}$-mesons will provide us very important information about
the heavy quark symmetry and the light quark dynamics.

The strong and radiative decays of the charmed mesons have been
studied in many different models. For example, the chiral
perturbation theory and heavy quark effect theory are used in
Refs.~\cite{Wise:1992hn,Burdman:1992gh,Yan:1992gz,Cheng:1992xi,Cho:1992nt,Amundson:1992yp,Casalbuoni:1996pg,Cheung:2014cka,Wang:2019mhm}.
Various quark models are employed in
Refs.~\cite{Godfrey:1985xj,Sucipto:1987qj,Kamal:1992uv,Barik:1994vd,Ivanov:1994ji,Jaus:1996np,Choi:2007se}.
There are also lots of other theoretical methods such as vector
meson dominance hypothesis~\cite{Colangelo:1993zq}, QCD sum
rules~\cite{Aliev:1994nq,Aliev:1995zlh,Dosch:1995kw,Zhu:1996qy,Wang:2015mxa},
quark-potential models
\cite{Godfrey:1985xj,Goity:2000dk,Ebert:2002xz,Simonis:2018rld},
extended Nambu-Jona-Lasinio model~\cite{Deng:2013uca,Luan:2015goa},
the cloudy bag model~\cite{Miller:1988tz}, the constituent
quark-meson model~\cite{Deandrea:1998uz}, lattice QCD
simulations~\cite{Becirevic:2009xp}, and so on.

For the ground states, the mass splittings between $D_{(s)}^\ast$
and $D_{(s)} $ just lie above the pion mass $m_\pi$ with $2-3$ MeV.
The constraint from phase space leads to the dominant pion and
photon emission decay modes of $D_{(s)}^*$, i.e. $D_{(s)}^*\to
D_{(s)}\gamma$ and $D_{(s)}^*\to D_{(s)}\pi$. For the charmed
strange meson $D_s^*$, the decay modes are particularly interesting.
$D_s^{\ast}\to D_s\pi^0$ is the strong decay process which violates
the isospin symmetry. The double suppressions from phase space and
the isospin violation make the hadron decay width tiny, at the order
of several eVs. The branch ratio of this strong decay mode is
$(5.8\pm0.7)\%$, which is much less than that of the electromagnetic
decay $D_s^*\to D_s\gamma$ about
$(93.5\pm0.7)\%$~\cite{Tanabashi:2018oca}. The decay mode challenges
our physical intuition about the magnitude of strong decay.

The decay ratio of $\Gamma(D_s^{*+}\to
D_s^{+}+\pi^0)/\Gamma(D_s^{*+}\to D_s^{+}+\gamma)$ have been
measured in CLEO~\cite{Gronberg:1995qp} and
BaBar~\cite{Aubert:2005ik}, respectively. Theoretically, this decay
channel has been studied in
Refs.~\cite{Cho:1994zu,Ivanov:1998wn,Terasaki:2015eao} with the
chiral symmetry and heavy quark symmetry, where only the tree level
contributions are considered. The very exotic hadronic decay mode
deserves more refined investigations.

The chiral perturbation theory is the effective field theory of low
energy QCD, which is a systematic and model-independent framework.
It is a powerful tool to analyze the physics associated with the
light degrees of freedom within the $D_{(s)}$-mesons below the
typical energy scale, $m_\rho$. For the $D_{(s)}$-mesons, the charm
quark mass $m_c$ is much larger than the light quark mass
$m_{q}~(q=u,d,s)$, thus $m_c$ can be integrated out at the low
energy scale. The color-magnetic interaction in the QCD Hamiltonian
is suppressed by $1/m_c$ and can be omitted at the leading order of
the heavy quark effective theory. Thus, heavy quark is regarded as
the static color source and the heavy quark spin symmetry is kept.

In
Refs.~\cite{Wise:1992hn,Yan:1992gz,Burdman:1992gh,Cheng:1993kp,Cheng:1993gc,Ivanov:1995yt,Ivanov:1997bh},
the chiral effective theory incorporating heavy quark symmetry was
constructed. In the effective theory, the chiral Lagrangian
describes the low energy strong interactions between the heavy
hadrons and light Goldstone bosons. Naturally, we can exploit this
chiral effective theory to describe strong decay of the
$D^*_{(s)}\to D_{(s)}\pi$.

 In this work, we focus on the isospin violating decay $D^*_s\to D_s\pi^0$.  We use the heavy meson chiral perturbation theory to investigate this process. Based on previous works, we not only calculate the leading order contribution, but also include the next-to-leading order loop diagrams and tree diagrams.
 The contributions of the loop diagrams manifest the complicated light quark dynamics, which generates some different structures from the leading ones. Besides, the $m_\pi$ dependent analytic expressions might be useful to do the extrapolations in lattice QCD simulations.

 This paper is organized as follows. In Sec.~\ref{sec_lag}, we give the effective Lagrangians with respect to the charmed mesons and light pseudoscalars. In Sec.~\ref{sec_decay}, we illustrate the Feynman diagrams of the decay $D^*_s \rightarrow D_s \pi^0$, the corresponding analytic expression of each diagram, and the numerical results, respectively. In Sec.~\ref{sec_dis}, we give some discussions and conclusions.

\section{Effective Lagrangians}\label{sec_lag}

One may use the chiral symmetry and the heavy quark symmetry to
construct the Lagrangians that account for the heavy mesons and
light pseudoscalars. The light pseudoscalar mesons octet are
described by the field $U(x)=u^2=e^{i\phi / f_\phi}$ with
 \begin{eqnarray}
    \phi=\left(\begin{array}{ccc}
    \pi^{0}+\frac{1}{\sqrt{3}}\eta & \sqrt{2}\pi^{+} & \sqrt{2}K^{+}\\
    \sqrt{2}\pi^{-} & -\pi^{0}+\frac{1}{\sqrt{3}}\eta & \sqrt{2}K^{0}\\
    \sqrt{2}K^{-} & \sqrt{2}\bar{K}^{0} & -\frac{2}{\sqrt{3}}\eta
    \end{array}\right),
 \end{eqnarray}
 and $f_\phi$ is the decay constants of the light pseudoscalars.
 Their experimental values are $f_\pi=92.4$ MeV, $f_K=113$ MeV and $f_\eta=116$ MeV, respectively.
 The chiral connection is defined as
 \begin{eqnarray}
 \Gamma_{\mu}\equiv\frac{1}{2}\left(u^{\dagger}\partial_{\mu}u+u\partial_{\mu}u^{\dagger}\right).
 \end{eqnarray}

 The leading order Lagrangian that describes the self-interaction of the octet pseudoscalars can be written as~\cite{Cho:1992gg,Cho:1994zu}
 \begin{eqnarray}\label{lag0}
 \mathcal{L}_{\phi\phi}=\frac{f_{\phi}^{2}}{4}\mathrm{Tr}\left[\partial_{\mu}U\partial^{\mu}U^{\dagger}\right]+\frac{f_{\phi}^{2}}{4}\mathrm{Tr}\left[\chi U^{\dagger}+U\chi^{\dagger}\right],
 \end{eqnarray}
 where $\mathrm{Tr}[\dots]$ denotes the trace in flavor space.
 The building block $\chi=2B_0m_q$ contains the light quark mass matrix $m_q$,
 \begin{equation}
    m_{q}=\left(\begin{array}{ccc}
    m_{u} & 0 & 0\\
    0 & m_{d} & 0\\
    0 & 0 & m_{s}
    \end{array}\right),
 \end{equation}
 and $B_0=-\langle \bar{q}q\rangle/(3f_\phi^2)$ is a parameter related to the quark condensate.
 The second term in Eq.~\eqref{lag0} embodies the chiral symmetry breaking effect, which implies the $\pi^0$ and $\eta$ mixing vertex, i.e.,
 \begin{equation}\label{eq_pi-eta}
 \mathcal{L}_{\mathrm{mixing}}=-\frac{B_0}{\sqrt{3}}(m_u-m_d)\eta\pi^0.
 \end{equation}
This equation demonstrates the origin of the isospin symmetry
violation at the quark level, i.e., the tiny mass difference between
$u$ and $d$ quarks.

The spin doublet of the anticharmed vectors $\bar{D}^{\ast}$ and
pseudoscalars $\bar{D}$ can be expressed as the four-velocity
dependent superfield $\mathcal{H}$ in the heavy quark limit, i.e.,
 \begin{eqnarray}
 \mathcal{H}&=&\left[P_{\alpha}^{*}\gamma^{\alpha}+iP\gamma_{5}\right]\frac{\left(1-v \!\!\!/\right)}{2},\nonumber
 \\
 \bar{\mathcal{H}}&=&\gamma_{0}\mathcal{H}^{\dagger}\gamma_{0}=\frac{1-v \!\!\!/}{2}\left[P_{\alpha}^{*\dagger}\gamma^{\alpha}+iP^{\dagger}\gamma_{5}\right],
 \end{eqnarray}
 where $v=(1,\bm{0})$ is the four-velocity of the heavy mesons, and the charmed meson fields are denoted as
 \begin{equation}
 P^{(*)}=(\bar{D}^{0(*)},D^{(*)-},D_s^{(*)-}).
 \end{equation}

 The leading order Lagrangian describing the low energy interactions of the anticharmed mesons and light pseudoscalars reads
 \begin{equation}\label{Lag1}
 \mathcal{L}_{P^{*}P\phi}^{(1)}=-i\langle\bar{\mathcal{H}}v\cdot \mathcal{D} \mathcal{H}\rangle-\frac{\Delta}{8} \langle\bar{\mathcal{H}}\sigma^{\mu\nu}\mathcal{H}\sigma_{\mu\nu}\rangle+g\langle\bar{\mathcal{H}}u \!\!\!/\gamma_{5}\mathcal{H}\rangle,
 \end{equation}
 where $\mathcal{D}_\mu=\partial_\mu+\Gamma_\mu$, and $\langle\dots\rangle$ denotes the trace in spinor space. $\Delta=m_{P^\ast}-m_P$ is the mass splitting between $\bar{D}^{\ast}$ and $\bar{D}$. $g\approx 0.59$ represents the axial coupling constant, which can be determined from the partial decay width of $D^{\ast+}\to D^0\pi^+$~\cite{Wang:2019mhm,Tanabashi:2018oca} or lattice QCD~\cite{Detmold:2012ge}.
 $u_\mu$ is the chiral axial-vector current, which reads
 \begin{equation}
 u_{\mu}\equiv\frac{i}{2}\left(u^{\dagger}\partial_{\mu}u-u\partial_{\mu}u^{\dagger}\right).
 \end{equation}
 In Eq.~\eqref{Lag1}, the first term describes the kinetic energy of the heavy mesons.
 The second term comes from the $1/m_Q$ correction of the next-to-leading order color-magnetic interaction in heavy quark expansion. The third term gives the coupling vertices of $\bar{D}^\ast\bar{D}\pi$ and $\bar{D}^\ast\bar{D}^\ast\pi$.

 Next we shall consider the contribution of the $\mathcal{O}(p^2)$ tree diagram.
 In order to construct such an $\mathcal{O}(p^2)$ Lagrangian to provide $D^*_s D_s \pi^0$ vertex, we need the building blocks $\chi_{-}$ and $\partial^{\mu}u_{\mu}$.
 If we use the building block $\chi_{-}$, one should notice that the parity of this building block is negative, i.e., we have to multiply another $\mathcal{O}(p^0)$ building block with negative parity to make sure the parity of the Lagrangian is positive.
 However, there does not exist such a building block that can satisfy both the requirement of parity conservation and Lorentz invariance.
 For the other building block $\partial^{\mu}u_{\mu}$, their exists the same problem.
 Thus, there does not exist $\mathcal{O}(p^2)$ chiral Lagrangian contributing to the isospin violating process after considering the constraint from Lorentz invariance and CPT conservation.

 In our calculation, we also consider the contribution from the loop diagrams, which will be presented in latter part. According to the power counting, the chiral order of the one-loop diagrams is at least $\mathcal{O}(p^3)$.
 In order to absorb the divergence in the loop diagrams, the $\mathcal{O}(p^3)$ tree-level Lagrangian is constructed as follows,
 \begin{eqnarray}\label{Lag2}
 \mathcal{L}_{P^{*}P\phi}^{(3)}&=&\frac{b_{1}}{\Lambda_\chi^{2}}\langle\bar{\mathcal{H}}u\!\!\!/\hat{\chi}_{+}\gamma_{5}\mathcal{H}\rangle+  \frac{b_{2}}{\Lambda_\chi^{2}}\langle\bar{\mathcal{H}}u\!\!\!/\gamma_{5}\mathcal{H}\rangle\mathrm{Tr}\left[\chi_{+}\right]\nonumber
 \\
 &&+i\frac{c_{1}}{\Lambda_\chi^{2}}\langle\bar{\mathcal{H}}\partial\!\!\!/\hat{\chi}_{-}\gamma_{5}\mathcal{H}\rangle+\frac{d}{\Lambda_\chi^{2}}\langle\bar{\mathcal{H}}\partial_{\nu}\partial\!\!\!/u^{\nu}\gamma_{5}\mathcal{H}\rangle\nonumber
 \\
&&+i\frac{c_{2}}{\Lambda_\chi^{2}}\langle\bar{\mathcal{H}}\gamma^{\mu}\gamma_{5}\mathcal{H}\rangle\partial_{\mu}\mathrm{Tr}\left[\chi_{-}\right],
 \end{eqnarray}
 where $\Lambda_\chi=4\pi f_\pi$. $b_1, b_2, c_1, c_2$ and $d$ are five low energy constants (LECs). The spurions $\chi_{\pm}$ are introduced as
\begin{eqnarray}
\chi_{\pm}&=&u^\dag\chi u^\dag\pm u\chi^\dag u,\quad
\hat{\chi}_\pm=\chi_\pm-\frac{1}{3}\mathrm{Tr}[\chi_\pm].
\end{eqnarray}

The Lagrangian~\eqref{Lag2} contains all possible relevant terms
satisfying the requirement of the symmetries. However, the
structures of the terms
$\langle\bar{\mathcal{H}}u\!\!\!/\gamma_{5}\mathcal{H}\rangle\mathrm{Tr}[\chi_{+}]$
and
$\langle\bar{\mathcal{H}}\gamma^{\mu}\gamma_{5}\mathcal{H}\rangle\partial_{\mu}\mathrm{Tr}[\chi_{-}]$
are the same as the ones from the leading order Lagrangian.
 Thus they can be absorbed into Eq.~\eqref{Lag1} by renormalizing the axial coupling $g$.
 The term $\langle\bar{\mathcal{H}}\partial_{\nu}\partial^{\nu}u_{\mu}\gamma^{\mu}\gamma_{5}\mathcal{H}\rangle$ is actually the same as the fourth term in the Lagrangian in our calculation, and we did not write it in Eq.~\eqref{Lag2}.
 With the above Lagrangians, we can analytically calculate the decay process $D^*_s\rightarrow D_s\pi^0$ up to $\mathcal{O}(p^3)$.

\section{Isospin Violating Decay}\label{sec_decay}

\subsection{Power counting and Feynman diagrams}\label{subsec_decay_dia}

 In chiral perturbation theory, one can use the power counting to assess the importance of Feynman diagrams generated by the effective Lagrangians when calculating the
 physical matrix element.
 The standard power counting for this process yields,
 \begin{eqnarray}\label{eq_power}
    \mathcal{O}=4N_L-2I_M-I_H+\sum_n nN_n,
 \end{eqnarray}
 where $N_L$, $I_M$ and $I_H$ are the numbers of loops, internal light pseudoscalar lines and internal heavy meson lines, respectively.
 $N_n$ is the number of vertices which are governed by the $n$-th order Lagrangians.
 Thus, we can write down the decay amplitude as the following expression,
 \begin{equation}
 \mathcal{M}=\mathcal{M}^{(1)}_{\mathrm{tree}}+\mathcal{M}^{(3)}_{\mathrm{tree}}+\mathcal{M}^{(3)}_{\mathrm{loop}},
 \end{equation}
 where the superscripts in the parentheses represent the chiral order.

\begin{figure}[hpt]
    \centering
    \includegraphics[width=0.25\textwidth]{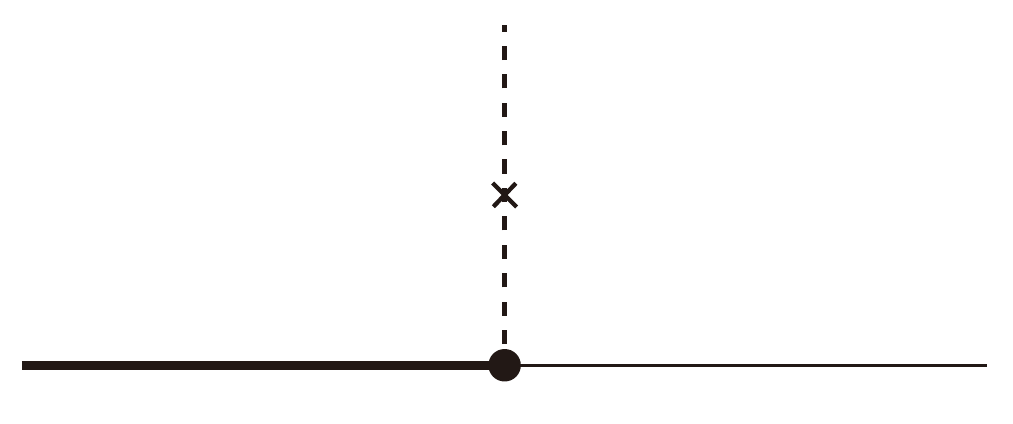}
    \caption{The tree diagram for the $D^{*}_{s}\rightarrow D_{s}\pi^0$ decay at the leading order. The thick solid, thin solid and dashed lines represent the heavy vector meson $D^*_s$, heavy pseudoscalar meson $D_s$, and light pseudoscalar mesons, respectively. The solid dot denotes the $\mathcal{O}(p)$ $D^\ast_s D_s\eta$ vertex, and the cross represents the $\eta-\pi$ mixing vertex.}
    \label{pictree}
\end{figure}

 For the $\mathcal{O}(p)$ tree diagram, the isospin violating effect comes from the $\pi-\eta$ mixing as shown in Fig.~\ref{pictree}. From Eq.~\eqref{eq_pi-eta}, the $\pi-\eta$ mixing effect comes from the mass difference between $u$ and $d$ quarks.

\begin{figure*}[hpt]
    \centering
    \includegraphics[width=0.85\textwidth]{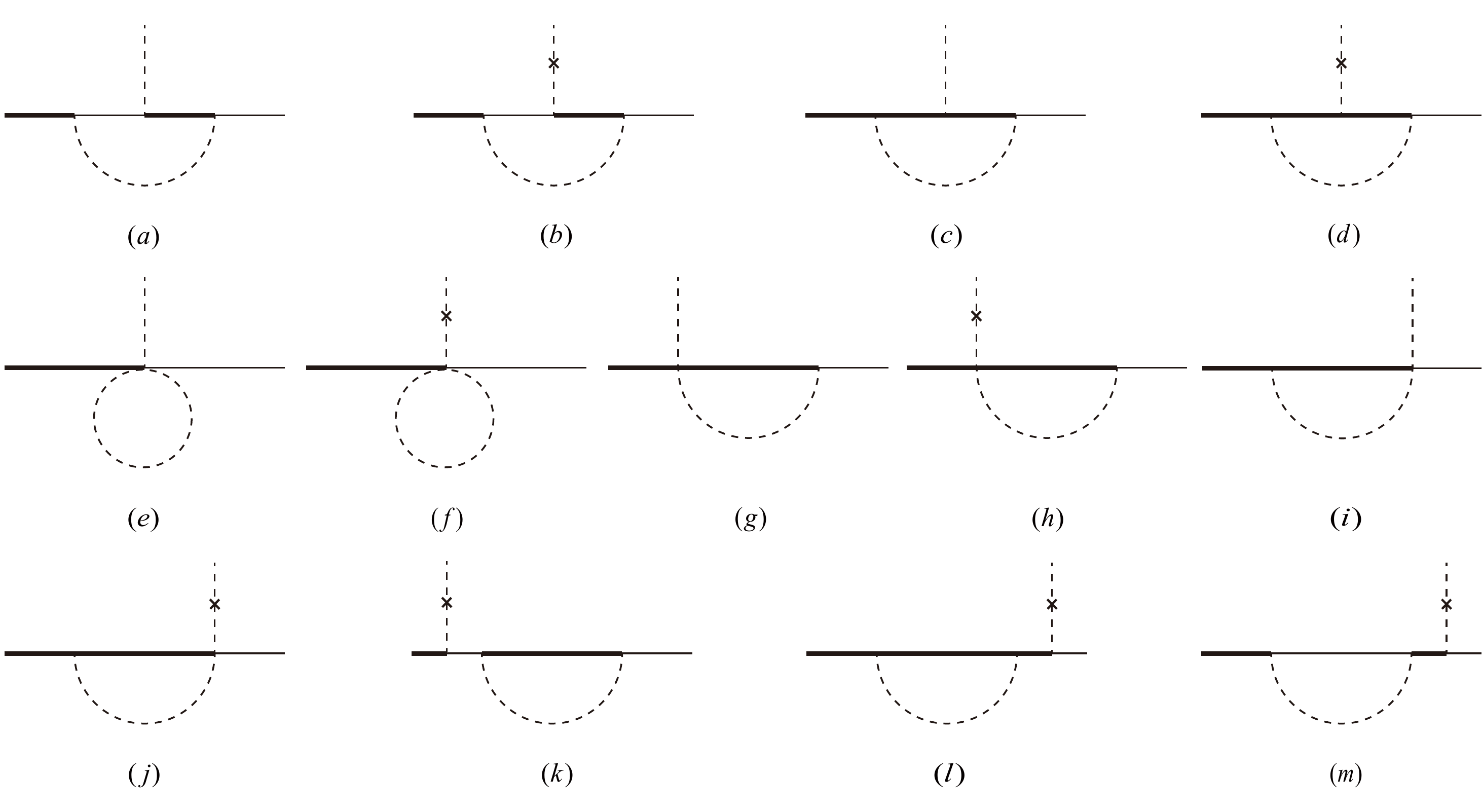}
    \caption{The loop diagrams for the $D^{*}_{s}\to D_{s}\pi^0$ decay at the next-to-leading order. The notations are the same as those in Fig. \ref{pictree}.}
    \label{picloop}
 \end{figure*}

 The loop diagrams with the vertices from the leading order Lagrangians [e.g., see Eqs.~\eqref{lag0},~\eqref{Lag1} and~\eqref{Lag2}] are shown in Fig. \ref{picloop}, which are $\mathcal{O}(p^3)$ diagrams according to the power counting law.
 The loop diagrams ($k$) and ($l$, $m$) are the renormalization of the $D_s$ and $D_s^*$ wave functions, respectively.
  \begin{figure}[hpt]
    \centering
    \includegraphics[width=0.25\textwidth]{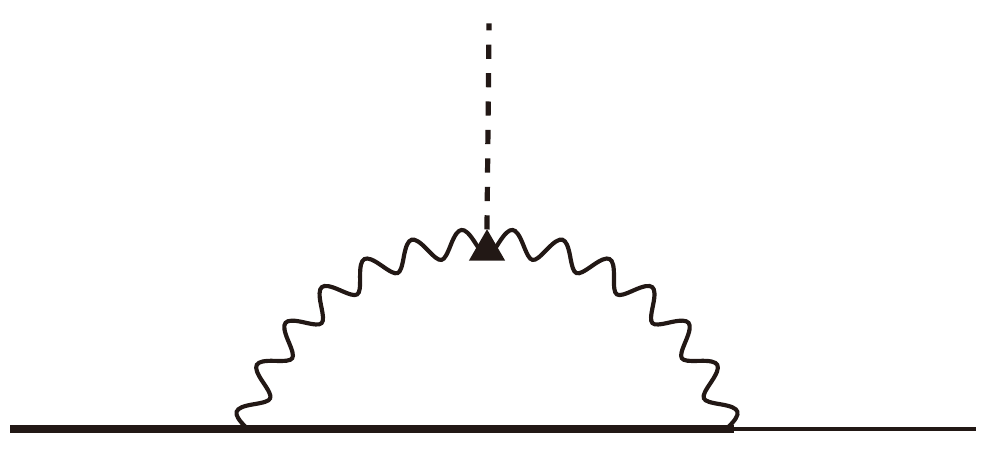}
    \caption{A diagrammatic presentation of the axial-vector current anomaly contribution to the  $D^{*}_{s}\to D_{s}\pi^0$ decay at the loop level. The wiggly line represents the photon, and the solid triangle denotes the $\pi^0\gamma\gamma$ coupling vertex. Other notations are the same as those in Fig. \ref{pictree}.}
    \label{pictree1}
 \end{figure}
  \begin{figure}[hpt]
    \centering
    \includegraphics[width=0.47\textwidth]{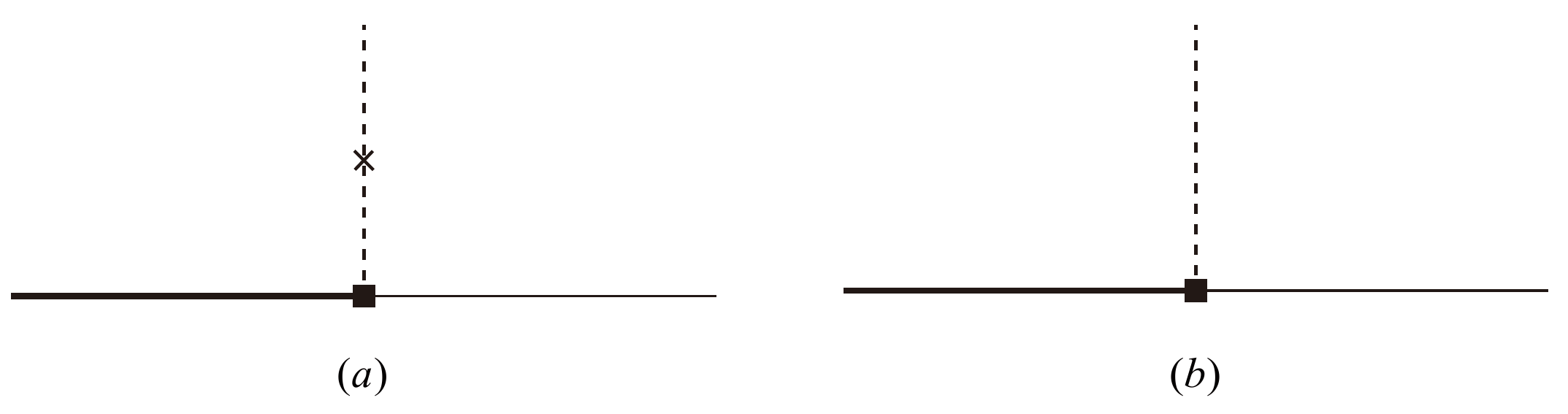}
    \caption{The tree diagrams for the $D^{*}_{s}\to D_{s}\pi^0$ decay at the next-to-leading order. The solid square stands for the $\mathcal{O}(p^3)$ coupling. Other notations are the same as those in Fig. \ref{pictree}.}
    \label{pictree3}
 \end{figure}

 The vertex with two heavy mesons and one light pseudoscalar comes from the third term of the $\mathcal{O}(p)$ Lagrangian~\eqref{Lag1}.
 The vertex denoted with the cross is from the Lagrangian~\eqref{eq_pi-eta}.
 The vertex in the diagram ($e$, $f$) connecting two heavy mesons and three pseudoscalars also stems from the third term of Eq.~\eqref{Lag1}, where we need to expand the axial-vector field $u_\mu$ to the second order.
 For the vertices with two heavy mesons and two light pseudoscalars in diagram ($g$, $h$, $i$, $j$), we can derive them in the first term of Eq.~\eqref{Lag1}.
 The chiral connection in the covariant derivative generates this kind of vertex.

 For the $\mathcal{O}(p^3)$ loop diagrams, the isospin violating effect comes from two processes.
 The graphs ($b$, $d$, $f$, $h$, $j$) contain the $\eta-\pi$ mixing vertex which resembles the $\mathcal{O}(p)$ tree diagram.
 For the second type of the loop diagrams ($a$, $c$, $e$, $g$, $i$), they do not have the direct isospin violating vertex, i.e., $\eta-\pi$ mixing.
The second type of isospin violation arises from incomplete
cancellation of diagrams considering the mass splitting of particles
within the same isospin multiplet in the loops.
 For example, we shall consider the internal light pseudoscalars such as $K^-$ and $\bar{K}^0$, when calculating the loop diagram ($a$).
 If we ignore the mass splitting between $K^-$ and $\bar{K}^0$, their contributions are exactly the same but with opposite sign.
 The graph ($a$) becomes nonvanishing and gives the isospin violating effect when the tiny mass difference $\delta_{m_K}=m_{\bar{K}^0}-m_{K^-}$ is kept.
 Actually, both types of isospin violating effects originate from the mass difference between the $u$ and $d$ quarks.

 Besides the mass splitting between $u$ and $d$ quarks, another source of the isospin violating effect stems from the electromagnetic interaction, the charge difference between $u$ and $d$ quarks. The Feynman diagram is shown in Fig.~\ref{pictree1}. The vertex $\pi^0\to 2\gamma$ denoted by the solid triangle arises from the axial-vector current anomaly.
 However, the Feynman amplitude of such a diagram is proportional to $\alpha^2$, where $\alpha$ is the fine structure constant. The contribution of this diagram is highly suppressed.
 Thus, it is reasonable to neglect the isospin violation from the electromagnetic interaction in our calculation.

 The tree diagrams with the vertices coming from the next-to-leading order Lagrangian~\eqref{Lag2} are also $\mathcal{O}(p^3)$. We show the diagrams in Fig.~\ref{pictree3}.
 The $\mathcal{O}(p^3)$ tree diagram can contain the $D_s^\ast D_s\pi^0$ vertex, which is different from the $\mathcal{O}(p)$ one.

\subsection{Analytical results}\label{subsec_decay_exp}
Using Eqs.~\eqref{Lag1} and Eq.~\eqref{eq_pi-eta}, one can easily
get the amplitude of the $\mathcal{O}(p)$ tree diagram [see Fig.
\ref{pictree}], which yields
 \begin{equation}\label{eqa_begin}
 i\mathcal{M}^{(1)}=-\frac{g}{f_\eta}(q\cdot\varepsilon)\frac{2}{3}\frac{m_{K^{0}}^{2}-m_{K^{+}}^{2}}{m_{\eta}^{2}-m_{\pi}^{2}},
 \end{equation}
 where $q$ and $\varepsilon$ are the momentum of $\pi^0$ and polarization vector of $D_s^\ast$, respectively.
 The parameter $B_0(m_d-m_u)$ in Eq.~\eqref{eq_pi-eta} has been replaced by $m_{K^{0}}^{2}-m_{K^{+}}^{2}$.

 The decay amplitudes of the $\mathcal{O}(p^3)$ loop diagrams in Fig. \ref{picloop} are given as follows,

 \begin{eqnarray}
 i\mathcal{M}^{(3)}_{(a)}&=&\frac{g^{3}}{2f_K^2f_\pi}(q\cdot\varepsilon)\bigg[-\frac{F\left(m_{K^{+}},\omega_{1},\delta_{1}\right)}{q_0+\Delta_1}\nonumber\\
&&+\frac{F\left(m_{K^{0}},\omega_{2},\delta_{2}\right)}{q_0+\Delta_2}\bigg],\\[0.3cm]
 i\mathcal{M}^{(3)}_{(b)}&=&\frac{g^{3}}{3f_\eta}(q\cdot\varepsilon)\frac{m_{K^{0}}^{2}-m_{K^{+}}^{2}}{m_{\pi}^{2}-m_{\eta}^{2}}\bigg[\frac{1}{2f_K^2}\frac{F\left(m_{K^{+}},\omega_{1},\delta_{1}\right)}{q_0+\Delta_1}\nonumber\\
 &&+\frac{1}{2f_K^2}\frac{F\left(m_{K^{0}},\omega_{2},\delta_{2}\right)}{q_0+\Delta_2}-\frac{2}{3f_\eta^2}\frac{F\left(m_{\eta},\omega_{3},\delta_{3}\right)}{q_0+\Delta_3}\bigg],\nonumber\\[0.3cm]\\
 i\mathcal{M}^{(3)}_{(c)}&=&\frac{g^{3}}{f_K^2f_\pi}(q\cdot\varepsilon)\bigg[\frac{F\left(m_{K^{+}},\omega_{1}-\Delta_{1},\delta_{1}\right)}{q_0}\nonumber\\
&&-\frac{F\left(m_{K^{0}},\omega_{2}-\Delta_{2},\delta_{2}\right)}{q_0}\bigg],\\[0.3cm]
 i\mathcal{M}^{(3)}_{(d)}&=&\frac{g^{3}(q\cdot\varepsilon)}{3f_\eta}\frac{m_{K^{0}}^{2}-m_{K^{+}}^{2}}{m_{\pi}^{2}-m_{\eta}^{2}}\bigg[\frac{F\left(m_{K^{+}},\omega_{1}-\Delta_{1},\delta_{1}\right)}{-q_0f_K^2}\nonumber\\
 &&-\frac{F\left(m_{K^{0}},\omega_{2}-\Delta_{2},\delta_{2}\right)}{q_0f_K^2}\nonumber\\
 &&+\frac{4}{3}\frac{F\left(m_{\eta},\omega_{3}-\Delta_{3},\delta_{3}\right)}{q_0f_\eta^2}\bigg],\\[0.3cm]
 i\mathcal{M}^{(3)}_{(e)}&=&\frac{g}{6f_K^2f_\pi}(q\cdot\varepsilon)\left[J_{0}^{c}\left(m_{K^{+}}\right)-J_{0}^{c}\left(m_{K^{0}}\right)\right],\\[0.3cm]
 i\mathcal{M}^{(3)}_{(f)}&=&\frac{g(q\cdot\varepsilon)}{6f_K^2f_\eta}\frac{m_{K^{+}}^{2}-m_{K^{0}}^{2}}{m_{\pi}^{2}-m_{\eta}^{2}}\left[J_{0}^{c}\left(m_{K^{0}}\right)+J_{0}^{c}\left(m_{K^{+}}\right)\right],\nonumber\\
 \\
 i\mathcal{M}^{(3)}_{(g)}&=&i\mathcal{M}^{(3)}_{(h)}=i\mathcal{M}^{(3)}_{(i)}=i\mathcal{M}^{(3)}_{(j)}=0.
 \end{eqnarray}

 For the renormalization of the wave functions of the $D_s$ meson,
 \begin{equation}
 i\mathcal{M}^{(3)}_{(k)}=i\mathcal{M}^{(1)}\delta Z_{D_s},
 \end{equation}
 where
 \begin{equation}\label{sigmaDs}
 \delta Z_{D_s}=Z_{D_s}-1=\frac{1}{2}\frac{\partial\Sigma_{D_S}(m_{\phi},\omega)}{\partial \omega}\Big|_{\omega=-\Delta_3}.
 \end{equation}
 And for the renormalization of the wave functions of the $D^*_s$ meson,
 \begin{equation}
 i\mathcal{M}^{(3)}_{(l+m)}=i\mathcal{M}^{(1)}\delta Z_{D_s^*},
 \end{equation}
 where
 \begin{equation}\label{sigmaDsast}
 \delta Z_{D_s^*}=Z_{D^{*}_s}-1=-\frac{1}{2}\frac{\partial\Sigma_{D_s^*}(m_{\phi},\omega,\delta)}{\partial\omega}\Big|^{\omega=\Delta_3}_{\delta=0}.
 \end{equation}
 In Eqs.~\eqref{sigmaDs} and~\eqref{sigmaDsast}, the expressions of $\Sigma_{D_s}$ and $\Sigma_{D_s^{*}}$ read,
 \begin{eqnarray}
 \Sigma_{D_s}&=&\left(1-d\right)g^{2}\left[\frac{2}{f_{K}^{2}}J_{22}^{a}\left(m_{K},\omega\right)+\frac{2}{3f_{\eta}^{2}}J_{22}^{a}\left(m_{\eta},\omega\right)\right],\nonumber\\[0.3cm]
 \Sigma_{D_s^{*}}&=&\frac{2g^{2}}{f_{K}^{2}}J_{22}^{A}\left(m_{K},\omega,\delta\right)+\frac{2g^{2}}{3f_{\eta}^{2}}J_{22}^{A}\left(m_{\eta},\omega,\delta\right),
 \end{eqnarray}
 where the functions $F(m,\omega,\delta)$, $J_c^0(m)$, and $J_{22}^a(m,\omega)$ are the loop integrals, which are calculated with the dimensional regularization in $d$ dimensions. Their definitions and expressions are collected in the Appendix~\ref{app_func}.
 $J_{22}^{A}$ is defined as
 \begin{eqnarray}
 J_{22}^{A}(m,\omega,\delta)&=&J_{22}^a(m,\omega)+2J_{22}^a(m,\delta).
\end{eqnarray}
 The parameters $\omega_{1,2,3}$, $\delta_{1,2,3}$ and $\Delta_{1,2,3}$ given as,
 \begin{align}
\omega_{1}&=E-m_{D^{0}},&\qquad
\omega_{2}&=E-m_{D^{-}},\nonumber\\
\omega_{3}&=E-m_{D_{s}},&\\
\delta_{1}&=E-q_0-m_{D^{0*}},&\qquad
\delta_{2}&=E-q_0-m_{D^{-*}},\nonumber\\
\delta_{3}&=E-q_0-m_{D_{s}^{*}},&\\
\Delta_{1}&=m_{D^{*0}}-m_{D^{0}},&\qquad
\Delta_{2}&=m_{D^{*-}}-m_{D^{-}},\nonumber\\
\Delta_{3}&=m_{D_{s}^{*}}-m_{D_{s}},&
 \end{align}
where $E$ is the energy of $D_s^\ast$, which equals to
$m_{D_s^\ast}$ in the center of mass frame of the initial state.

 For the $\mathcal{O}(p^3)$ tree diagrams in Fig.~\ref{pictree3}, their amplitudes read,
 \begin{equation}
 i\mathcal{M}_{\mathrm{tree}}^{(3)}=i\mathcal{M}_{(a1)}+i\mathcal{M}_{(a2)}+i\mathcal{M}_{(b)},
 \end{equation}
 with
 \begin{eqnarray}
 i\mathcal{M}_{(a1)}&=&i\mathcal{M}^{(1)}\frac{1}{g\Lambda_\chi^{2}}\bigg[2\left(b_{1}-2c_{1}\right)m_{\eta}^{2}-\left(2b_{1}+d_{1}\right)m_{\pi}^{2}\bigg],\nonumber \\
 i\mathcal{M}_{(a2)}&=&i\mathcal{M}^{(1)}\frac{1}{g\Lambda_\chi^{2}}\bigg[3\left(b_{2}-2c_{2}\right)m_{\eta}^{2}+3\left(b_{2}+2c_{2}\right)m_{\pi}^{2}\bigg],\nonumber \\
 i\mathcal{M}_{(b)}&=&i\mathcal{M}^{(1)}\frac{1}{g\Lambda_\chi^{2}}\left(4c_{1}-6c_{2}\right)\left(m_{\eta}^{2}-m_{\pi}^{2}\right),\label{eqa_end}
 \end{eqnarray}
 where $\mathcal{M}^{(1)}$ is the $\mathcal{O}(p)$ amplitude in Eq.~\eqref{eqa_begin}.
 The contribution of the first $\mathcal{O}(p^3)$ tree diagram contains two parts, $i\mathcal{M}_{(a1)}$ and $i\mathcal{M}_{(a2)}$.
 The second part can be absorbed into the leading order diagram, because they have the same Lorentz structure except a constant factor.
 We ignore the isospin breaking effect from the decay constants of the light pseudoscalar mesons when calculating the contribution of the loop diagrams.
 Because the isospin breaking effect from the $K$ meson decay constant is about $0.1\%$~\cite{Cirigliano:2011tm,Carrasco:2014poa,Tanabashi:2018oca}.

 After performing the average over the initial $D_s^\ast$ polarization, the decay width of $D_s^\ast\to D_s\pi^0$ can then be written as
\begin{eqnarray}
\Gamma[D_s^\ast\to
D_s\pi^0]=\frac{|\bm{q}|^3}{24\pi}\frac{m_{D_s}}{m_{D_s^\ast}}|\mathcal{M}|^2.
\end{eqnarray}
\subsection{Numerical results}\label{sec_dis}
 We have derived the analytical expressions of the isospin violating decay $D_s^*\to D_s\pi^0$ with the chiral perturbation theory up to $\mathcal{O}(p^3)$.
 However, the $\mathcal{O}(p^3)$ Lagrangian [see Eq.~\eqref{Lag2}] contains unknown LECs, which are hard to be determined at present.
 In order to include the effects of the $\mathcal{O}(p^3)$ tree diagrams, we use two different strategies to estimate their contributions.

 {\it Strategy A}:
 We first adopt the nonanalytic dominance approximation~\cite{Bijnens:1995yn,Liu:2012uw,Wang:2018atz} to estimate the $\mathcal{O}(p^3)$ tree diagram contributions.
 We know that in the chiral perturbation theory, the amplitude of a tree diagram is the polynomials of $m_\phi^2$ and $q^2$, i.e., it only contains the analytic terms.
 While for a loop diagram, its amplitude might not only contain the polynomials of $m_\phi^2$ and $q^2$, but also have the typical multivalued functions, such as logarithmic and square root terms, which are called as the nonanalytic terms.
 The nonanalytic dominance approximation assumes that the analytic part of $\mathcal{O}(p^3)$ loop diagrams and the $\mathcal{O}(p^3)$ tree diagrams are roughly the same.
 This approximation might be rough to some extent, but can give us some clear indications about the convergence of the chiral expansion.

We then use this strategy to estimate the $\mathcal{O}(p^3)$ tree
level contribution and treat it as the error of our numerical
result. Our calculation yields
 \begin{eqnarray}\label{GammaDsastDspi}
 \Gamma[D_s^\ast\to D_s\pi^0]=(3.38\pm 0.12)~\text{eV}.
 \end{eqnarray}
Considering the $\Gamma[D_s^\ast\to
D_s\pi^0]/\Gamma[D_s^\ast]=(5.8\pm0.7)\%$, we can estimate the total
width of $D_s^\ast$ with the value in Eq.~\eqref{GammaDsastDspi},
 \begin{eqnarray}\label{GammaDsast}
  \Gamma[D_s^\ast]=58.26^{+10.37}_{-8.11}~\text{eV}.
 \end{eqnarray}

 The contributions are listed in Table~\ref{table_re1} order by order.
 The results are given in the cases of $\Delta\neq0$ and $\Delta=0$, respectively, where $\Delta=m_{D_{(s)}^\ast}-m_{D_{(s)}}$.
 For example, for the case of $\Delta\neq0$, we keep all the physical mass splittings in the loops.
 While for the case of $\Delta=0$, i.e., in the heavy quark limit, we neglect the mass difference of $D_{(s)}^\ast$ and $D_{(s)}$.

 From Table~\ref{table_re1}, we see that the variation of the total decay width of $D_s^\ast\to D_s\pi^0$ is not obvious, whereas the change of contribution from the $\mathcal{O}(p^3)$ loop diagrams is dramatic with $\Delta\neq0$ and $\Delta=0$.
 In other words, the heavy quark symmetry breaking effect at the loop level is very significant for the charm sectors.
 This effect has been noticed by some previous works~\cite{Wang:2019mhm,Wang:2019ato}.
 Additionally, we give the contributions of each $\mathcal{O}(p^3)$ loop diagram in Table~\ref{table_re2}.
 We also notice that the convergence of the chiral expansion is very good, even if we work in the SU(3) case.
 The convergence of the $\Delta\neq0$ case is much better than that of the $\Delta=0$ case.
 In Eqs.~\eqref{GammaDsastDspi} and~\eqref{GammaDsast} we adopt the $\Delta\neq0$ result to predict the decay width and total width of $D_s^\ast$.
 \begin{table*}
 \caption{The contributions order by order and the decay width of $D_s^\ast\to D_s\pi^0$ with $\Delta\neq0$ and $\Delta=0$, respectively. We give the numerical results of the structure $i\mathcal{M}/(q\cdot\epsilon)$ in unit of $10^{-3}\text{GeV}^{-1}$, and the decay width in unit of eV.}\label{table_re1}
 \setlength{\tabcolsep}{3.8mm}
  \begin{tabular}{c|cccc|c}
    \hline \hline
    Mass splitting & $\mathcal{O}(p)$ & $\mathcal{O}(p^{3})_{\mathrm{loop}}$ & $\mathcal{O}(p^{3})_{\mathrm{tree}}$ & Total & $\Gamma[D_s^\ast\to D_s\pi^0]$\\
    \hline
    $\Delta\neq0$ & $-46.90$ & $-2.73$ & $\pm1.08$ & $-(49.63\pm1.08)$ & $(3.38\pm0.12)$ eV\\
    $\Delta=0$ & $-46.90$ & $-14.69$ & $\dots$ & $-61.59$ & $5.20$ eV\\
    \hline \hline
  \end{tabular}
 \end{table*}

 \begin{table*}
 \caption{The contributions of each $\mathcal{O}(p^3)$ loop diagram with $\Delta\neq0$ and $\Delta=0$, respectively. We give the numerical results of the structure $i\mathcal{M}/(q\cdot\epsilon)$ in unit of $10^{-3}\text{GeV}^{-1}$.}\label{table_re2}
 \setlength{\tabcolsep}{3.8mm}
  \begin{tabular}{c|cccccccc}
    \hline \hline
    mass spliting & $a$ & $b$ & $c$ & $d$ & $e$ & $f$ & $k$ & $l+m$\tabularnewline
    \hline
    $\Delta\neq0$ & -1.94 & -1.65 & 7.15 & 3.12 & 1.45 & -4.86 & 3.03 & -9.03\tabularnewline
    $\Delta=0$ & 0.64 & -0.25 & -1.27 & 0.51 & 1.45 & -4.86 & 4.06 & -6.83\tabularnewline
    \hline \hline
  \end{tabular}
 \end{table*}

 {\it Strategy B}:
 We consider the naturalness of the chiral perturbation theory~\cite{Epelbaum:2008ga,Meng:2018zbl}.
 The amplitude can be expanded generally in power series of $q/\Lambda_\chi$ as follows,
 \begin{equation}
 \mathcal{M}=\mathcal{M}^{(0)}\underset{\mu}{\sum}\left(\frac{q}{\Lambda_{\chi}}\right)^{\mu}\mathcal{F}(g_{i}),
 \end{equation}
 where $\mathcal{M}^{(0)}$ is the leading order amplitude, $\mu$ is the chiral order, and $\mathcal{F}(g_{i})$ is a function of LECs.
 Therefore, in order to keep the convergence of the chiral expansion, a natural assumption requires the function $\mathcal{F}(g_{i})$ should be order one.
 The above is the naturalness assumption of the chiral perturbation theory.

\begin{figure}[hpt]
    \centering
    \includegraphics[width=0.5\textwidth]{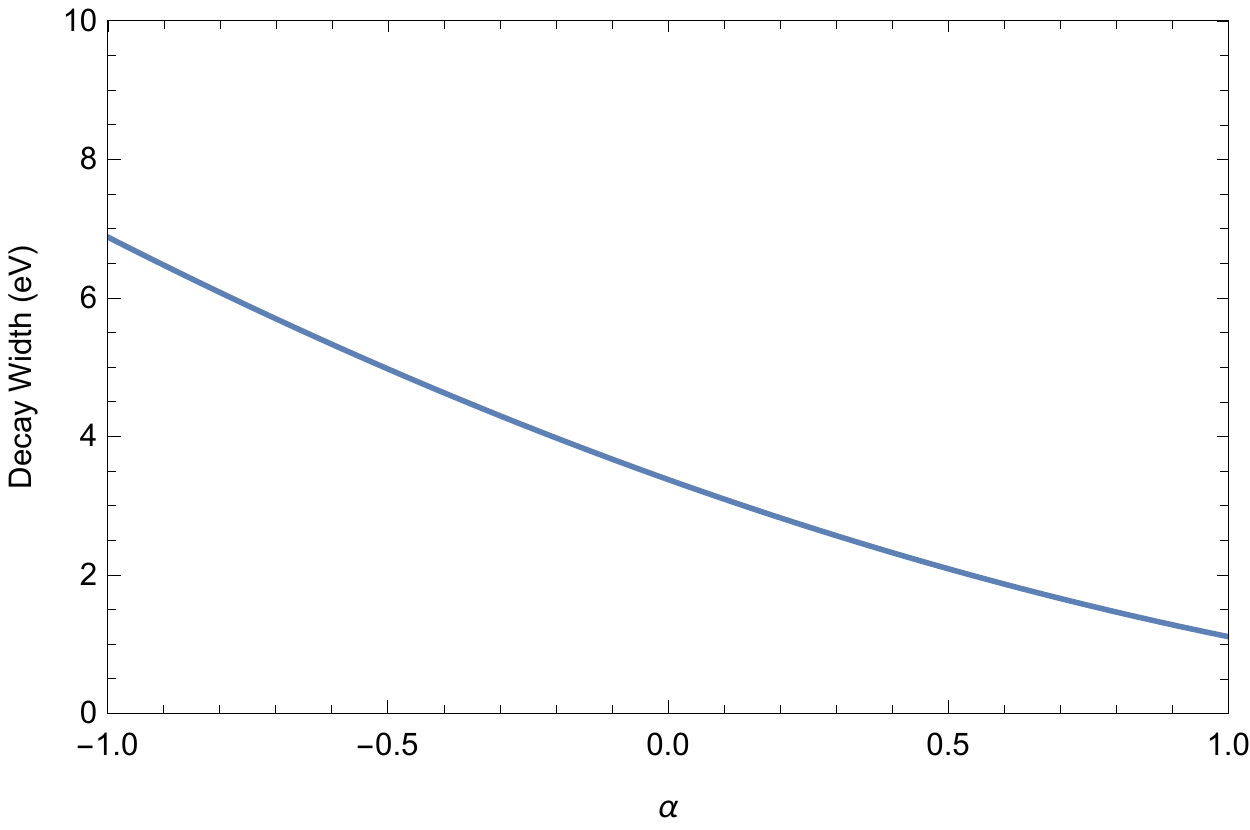}
    \caption{The change of the decay width of $D^*_s\to D_s\pi^0$ with the parameter $\alpha$.}
    \label{decay_width}
 \end{figure}

 For the $\mathcal{O}(p^3)$ tree diagrams with unknown LECs, except the terms which can be absorbed by $\mathcal{O}(p^1)$ Lagrangian, we can rewrite the remaining two parts as follows,
 \begin{equation}\label{imtree3a1}
 i\mathcal{M}_{tree}^{(3)a1}=i\mathcal{M}^{(1)}\frac{1}{\left(4\pi F_{\pi}\right)^{2}}\alpha\left(-m_{\eta}^{2}-\frac{3}{2} m_{\pi}^{2}\right),
 \end{equation}
 \begin{equation}\label{imtree3b}
 i\mathcal{M}_{tree}^{(3)b}=i\mathcal{M}^{(1)}\frac{1}{\left(4\pi F_{\pi}\right)^{2}}\alpha\left(-m_{\eta}^{2}+m_{\pi}^{2}\right).
 \end{equation}
 Here we replace all the $\mathcal{O}(p^3)$ LECs as "$\alpha g/2$", where $g$ is the LEC of the leading order Lagrangian, and parameter $\alpha$ is a order one number.
 The effect of the $\mathcal{O}(p^3)$ LECs can be roughly represented by the size of the parameter $\alpha$.
 Thus, in order to discuss the contribution of the $\mathcal{O}(p^3)$ tree diagrams as much as possible, we change the parameter from -1 to 1.
 The change of the total decay width with the parameter is shown in Fig.~\ref{decay_width}.
 When the $\alpha$ varies from from -1 to 1, the total decay is 1.11-6.88 eV.
 We can see that the contribution of the $\mathcal{O}(p^3)$ tree diagrams could be quite large.
 Nominally, the $\mathcal{O}(p^3)$ tree diagrams should be suppressed by the factor  $1/\left(4\pi F_{\pi}\right)^2$.
 But the $\eta$ meson mass is $547.8$ MeV, which makes the correction not as small as one naively guesses.
 Thus, the $\mathcal{O}(p^3)$ correction is important.

\section{Summary}\label{sec_dis}
 The heavy quark spin symmetry implies that the mass difference between the vector mesons $D^*$ and pseudoscalar mesons $D$ is small.
 Their mass splittings just lie above the pion mass with $2-3$ MeV.
 Therefore, the lowest $D^*$ mesons only have two main decay modes.
 One is the pion emission strong decay $D^*\rightarrow D\pi$, and the other one is the electromagnetic $D^*\rightarrow D\gamma$ decay.
 Generally, the decay width of the later one is usually much smaller than the first one due to the strength of the interactions. However, for the charmed strange meson $D_s^*$, the strong decay mode $D_s^\ast\to D_s\pi^0$ is much smaller than the electromagnetic one ~\cite{Tanabashi:2018oca} due to the double suppression of the phase space and isospin violation.

 In this work, we have systematically calculated the isospin violating decay $D_s^*\to D_s\pi^0$ with the heavy meson chiral perturbation theory up to the $\mathcal{O}(p^3)$ including the loop diagrams.
 The analytical expressions are derived up to chiral order $\mathcal{O}(p^3)$.
 For this process, the $\mathcal{O}(p^2)$ Lagrangian does not exist under constraint of the parity and Lorentz symmetries.
 The corrections to the leading order contribution come from the $\mathcal{O}(p^3)$ tree and loop diagrams.
 The vertices of the $\mathcal{O}(p^3)$ loop diagrams are governed by the leading order Lagrangians.
 Thus, the numerical result of the loop diagrams only depends on one parameter $g$, which has been well determined by experiments and lattice QCD.
 Our calculation of the leading order amplitude and $\mathcal{O}(p^3)$ loop diagrams shows very good convergence of the chiral expansion.
 The convergence in the $\Delta\neq0$ case is much better than that in the $\Delta=0$ one.

 The $\mathcal{O}(p^3)$ tree level amplitudes contain four undetermined LECs.
 We use two strategies to estimate the uncertainty of the $\mathcal{O}(p^3)$ tree level contributions.
 With the nonanalytic dominance approximation, we get the $\Gamma[D_s^\ast\to D_s\pi^0]=(3.38\pm 0.12)$~eV.
 With the naturalness assumption of the chiral perturbation theory, we give a possible range of the isospin violating decay width, $[1.11, 6.88]$~eV.
 We find that the contribution of the $\mathcal{O}(p^3)$ tree diagrams might be significant compared with the leading order one.

 The isospin violating decay plays a very important role in studying the character and structure of the $D_s^\ast$ meson.
 We expect experiments and lattice QCD can provide more results about the decays of the charmed mesons in the future.
 Our analytical expressions can also be helpful to the chiral extrapolations in lattice QCD simulations.

\section*{ACKNOWLEDGEMENTS}

B. Yang is very grateful to W. Z. Deng,  X. L. Chen for very helpful
discussions. This project is supported by the National Natural
Science Foundation of China under Grant 11975033.

\begin{appendix}

\section{Definitions and expressions of the loop integrals} \label{app_func}
 The loop functions used in Eqs.~\eqref{eqa_begin}-\eqref{eqa_end} are defined as follows,
 \begin{eqnarray}
 F\left(m_{\phi},\omega,\delta\right)&\equiv&\frac{1}{d-1}\bigg[(m_{\phi}^{2}-\delta^{2})J^{a}_{0}\left(m_{\phi},\delta\right)-(m_{\phi}^{2}-\omega^{2})\nonumber\\
 &&\times J^{a}_{0}\left(m_{\phi},\omega\right)+\left(\delta-\omega\right)J^{c}_{0}\left(m_{\phi}\right)\bigg],
 \end{eqnarray}
 \begin{equation}
 J^{c}_{0}(m_{\phi})\equiv i\int\frac{d^{d}k\lambda^{4-d}}{\left(2\pi\right)^{d}}\frac{1}{k^{2}-m_{\phi}^{2}+i\epsilon},
 \end{equation}
 \begin{equation}
 J^{a}_{0}\left(m_{\phi},\omega\right)\equiv i\int\frac{d^{d}k\lambda^{4-d}}{\left(2\pi\right)^{d}}\frac{1}{\left[k^{2}-m_{\phi}^{2}+i\epsilon\right]\left[v\cdot k+\omega+i\epsilon\right]},
 \end{equation}
 \begin{eqnarray}
 &&i\int\frac{d^{d}k\lambda^{4-d}}{\left(2\pi\right)^{d}}\frac{k^{\mu}k^{\nu}}{\left[k^{2}-m_{\phi}^{2}+i\epsilon\right]\left[v\cdot k+\omega+i\epsilon\right]}\nonumber\\
 &&\qquad\equiv v^{\mu}v^{\nu}J^{a}_{21}\left(m_{\phi},\omega\right)+g^{\mu\nu}J^{a}_{22}\left(m_{\phi},\omega\right),
 \end{eqnarray}

 The above loop integrals can be calculated with the dimensional regularization in $d$ dimensions. Their expressions read
 \begin{equation}
 J^{c}_{0}(m)=-\frac{m^{2}}{16\pi^{2}}\left(L+\text{ln}\frac{\lambda^{2}}{m^{2}}\right),
 \end{equation}
  \begin{equation}
 J_{22}^{a}\left(m,\omega\right)=\frac{1}{d-1}\left[\left(m^{2}-\omega^{2}\right)J^{a}_{0}(m,\omega)+\omega J^{c}_{0}(m)\right].
 \end{equation}
 We adopt the $\overline{\text{MS}}$ scheme to renormalize the loop integrals.
 The $L$ is defined as follows,
 \begin{equation}
L=\frac{2}{4-d}+\text{ln}4\pi-\gamma_{E}+1,
 \end{equation}
 where $\gamma_{E}\approx0.5772$ is the Euler-Mascheroni constant.
 \begin{widetext}
 \begin{equation}
 J^{a}_{0}\left(m,\omega\right)=\begin{cases}
-\frac{\omega}{8\pi^{2}}(L+\text{ln}\frac{\lambda^{2}}{m^{2}}+1)+\frac{1}{4\pi^{2}}\sqrt{\omega^{2}-m^{2}}\text{arccosh}\left(\frac{\omega}{m}\right)-\frac{i}{4\pi}\sqrt{\omega^{2}-m^{2}} & (\omega>m)\\
-\frac{\omega}{8\pi^{2}}(L+\text{ln}\frac{\lambda^{2}}{m^{2}}+1)+\frac{1}{4\pi^{2}}\sqrt{m^{2}-\omega^{2}}\text{arccos}\left(-\frac{\omega}{m}\right) & (-m<\omega<m)\\
-\frac{\omega}{8\pi^{2}}(L+\text{ln}\frac{\lambda^{2}}{m^{2}}+1)-\frac{1}{4\pi^{2}}\sqrt{\omega^{2}-m^{2}}\text{arccosh}\left(-\frac{\omega}{m}\right)
& (\omega<-m)
\end{cases}.
 \end{equation}
 \end{widetext}
\end{appendix}

\vfil \thispagestyle{empty}
\newpage

\bibliography{ref}

\end{document}